\documentclass[%
 reprint,
 amsmath,amssymb,
 aps,
pra,
]{revtex4-2}

\usepackage{graphicx}
\usepackage{hyperref}
\usepackage{dcolumn}
\usepackage{bm}
\usepackage{graphicx}
\usepackage{mwe}
\usepackage{float}
\usepackage{tabularx}
\usepackage{booktabs}
\usepackage{siunitx}
\usepackage{graphicx}

\begin{document}

\preprint{APS/123-QED}

\title{Monochromation of pulsed electron beams with terahertz radiation at a planar mirror}

\author{Cecilia Abbamonte}
\author{Adam Bartnik}
\author{Jared Maxson}
\affiliation{Cornell Laboratory for Accelerator-Based Sciences and Education, Ithaca, NY }%

\date{\today}

\begin{abstract}
Exquisite control of electron beam energy is required for many electron spectroscopy and imaging
applications. For both continuous and pulsed beams, the beam energy spread is fundamentally limited by the electron source, and is typically a sizable fraction of an electron-volt. In this paper, we present a means to reduce electron beam energy spread after emission to the level of a few 10s of meV rms using femtosecond photoemission and an interaction with laser-derived single- to few-cycle terahertz (THz) radiation.  We show analytically and in particle tracking simulations that this interaction can remove energy spread stored in both the transverse and longitudinal degrees of freedom. We analytically formulate the limit of energy spread that this technique can achieve, and map the non-ideal affects arising at high frequencies. The interaction is mediated by the beam's passage through a mirror which is reflective to terahertz radiation but allows transmission of the majority of the electron beam (e.g. a wire mesh). This method then only requires beam current losses of a few tens of percent, far smaller than what is achieved in prism and slit-based electron monochromators. \\

\end{abstract}

\maketitle


\section{Introduction}

Femtosecond photoemission has been widely exploited to generate extremely short pulses of electrons for use in microscopy and scattering experiments, providing temporal resolution sufficient to resolve atomic motion in states far from equilibrium \cite{Zewail2006, Sciaini2011, Barwick2009, Filippetto2022}. A distinguishing feature of pulsed electron beams is that they are manipulatable by time varying electromagnetic fields, granting control over the energy-time phase space of the electron pulse. Radiofrequency (rf) cavities are frequently used to improve time resolution in ultrafast experiments by creating an energy-time correlation that compresses the duration of the electron pulse \cite{Chatelain2012, vanOudheusden2010, Filippetto2022}, and other methods of pulse compression that take advantage of energy-time correlation with dispersive magnetic optics are also under active development
\cite{Qi2020, yang2025}. In this work our focus is on the conjugate process to compression: exploiting the time-space-energy correlations that naturally arise in beam transport for precise control of the beam’s energy distribution. 

Electron beams with narrow energy spreads are critical for continuous and time resolved experiments across a wide range of spectroscopic and imaging applications. In particular, electron energy loss spectroscopy (EELS) in electron microscopes plays an important role in probing the electronic degrees of freedom \cite{Abbamonte2025, Sato2011, Terauchi1991}, identifying chemical composition \cite{Colliex2019, Hachtel2019}, measuring the local bonding environment of individual atoms \cite{Hage2020, Dwyer2016, Ramasse2013}, and with extremely sharp energy resolution (few 10s of meV and less), EELS can also be sensitive to phonon populations \cite{Gadre2022, Yan2021}. To reach these narrow energy widths, typically electrons are dispersed by a prism (dipole bending magnet) and collimated with a slit, which imposes a heavy loss in beam current and often preserves only a few percent of the incident beam \cite{Krivanek2014}. For this reason, ultrafast EELS, with a pulsed and inherently lower current electron beam, has not typically employed this type of monochromation \cite{Pomarico2017, Kim2023, Madan2019, Carbone2009, Piazza2013}. Monochromation that largely preserves beam current is of critical importance for obtaining energy resolution in the ultrafast case that is comparable to that of a continuous beam. Finally, we note that monochromatized electron beams do not suffer from chromatic aberrations and have long temporal coherence and thus may drive progress in conventional \cite{Freitag2005, Mihaila2025} and quantum electron optics in general \cite{Ruimy2025}. 

A promising alternative to lossy prism-based monochromators involves using electromagnetic waves to reduce energy spread by manipulating the phase space of the electron beam. A design proposed in Ref. \cite{Duncan2020} uses radiofrequency (rf) cavities in two steps to losslessly correct energy spread correlated with  both the temporal (longitudinal) and spatial (transverse) beam coordinates. A subsequent experiment described in Ref.~\cite{Yannai2023} demonstrates several key features of this design, in which a terahertz (THz) frequency pulse is shown to losslessly reduce the energy spread in the time domain by a factor of 3. THz radiation has also proved to be a useful tool in the control of electron beam phase space for other applications including bunching \cite{Kealhofer2016} \cite{Curry2018}, acceleration \cite{Nanni2015}, and time of flight energy analysis \cite{Ehberger2018-EEA} for characterization of the temporal beam profile. Advantageously, THz radiation derived from laser sources (for example by optical rectification) is intrinsically synchronized to the electron source, thereby eliminating a dominant source of phase jitter by avoiding the need for phase synchronizing electronics. 

\begin{figure*}
    \centering
    \includegraphics[width=\linewidth]{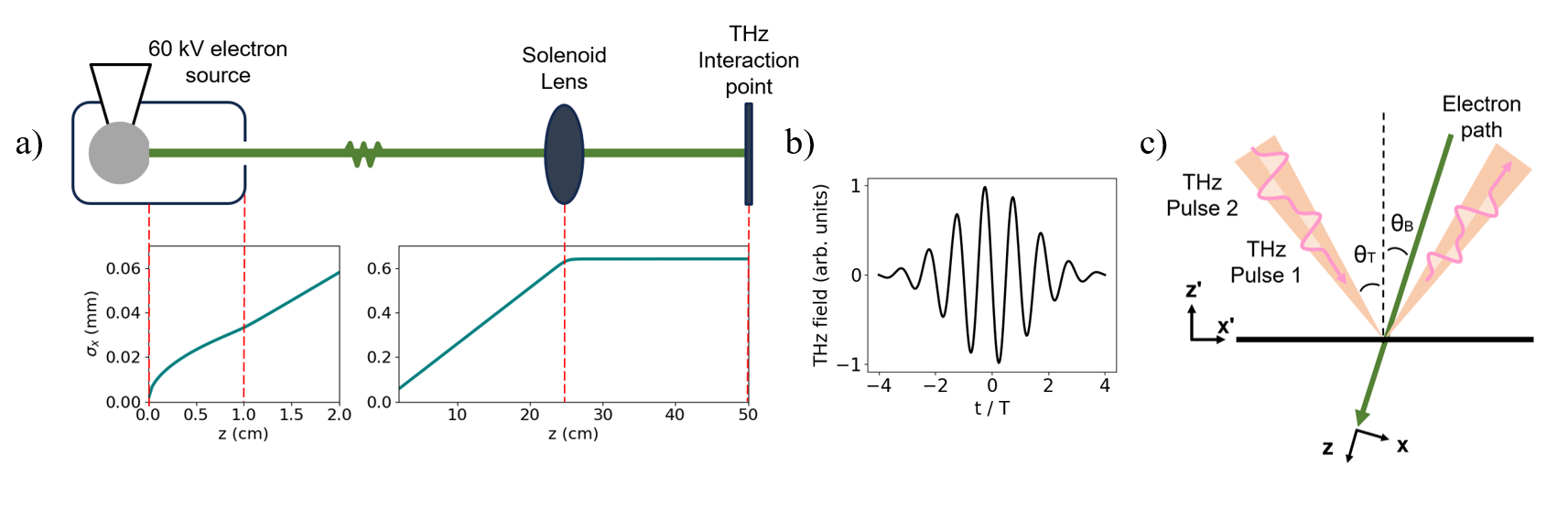}
    \caption{Schematics of the THz monochromator and beamline. $\boldsymbol{a)}$ Electrons are emitted from a flat photocathode and accelerated by a 60 kV dc gun with a peak field of 6 MV/m and an anode diameter of 3 mm. The beam is collimated in a solenoid before passing through the THz mirror monochromator.
    $\boldsymbol{b)}$ The THz pulse shape used for this set of simulations, where $t$ is the time of arrival and $T$ is the period of one cycle. $\boldsymbol{c)}$ The two THz pulses reflect off of a mirror at angle $\theta_T$ to the normal. The pulses are P-polarized, with $\vec{E}$ in the $x$-$z$ plane and $\vec{B}$ along $\hat{y}$. An electron with zero transverse momentum passes through the mirror at angle $\theta_B$.}
    \label{fig:schematics}
\end{figure*}

This work builds upon the rf monochromator design of \cite{Duncan2020} and the THz monochromation in a working electron microscope demonstrated in \cite{Yannai2023}. We propose a monochromation scheme that uses two THz pulses to correct energy spread stored in the longitudinal and transverse degrees of freedom. Interaction between the electrons and a THz wave is facilitated by a mirror which is reflective to THz radiation but largely transmissive to the electron beam, for example a thin mirror membrane or a wire mesh TEM grid. The mechanism for this interaction is well-established both in theory and experimentally \cite{Kealhofer2016, Ehberger2018-tilted, Ehberger2018-EEA, Morimoto2018}. Unlike \cite{Duncan2020}, monochromation with the THz mirror is done at a single location (with one or two pulses) along the electron beam's trajectory, and as we will show, is not likely to be dominated in practice by phase and amplitude jitter. The major downsides to this approach compared to the rf monochromator are electron loss at the mirror and slightly worse monochromation performance as compared to the ideal rf case due to the shorter period of the THz pulse, as described next. In contrast to \cite{Yannai2023}, this method corrects energy spread correlated with both the transverse and longitudinal degrees of freedom, which provides more substantial monochromation as compared to correcting using the energy-time correlation alone. We also choose electron optics that enhance the energy-time-space correlation, which is straightforward with a custom electron source.


The fundamental principle at work is that an electron emitted from a photocathode due to a laser pulse with vanishing duration and transverse size will have a downstream position and arrival time perfectly correlated with its energy: higher energy electrons arrive earlier and tend to have larger transverse radii. This energy-space correlation is parabolic in time and transverse radius. By using two THz pulses with Gaussian transverse distributions, we determine a condition in which the THz-driven energy gain of the electron beam as a function of position and arrival time matches this paraboloid to good approximation. The degree of monochromation is then limited by two effects: 1) the short THz period limits the temporal duration of the electron beam that can be monochromated, 
2) finite emission size and duration partially scrambles the perfect energy-time-radius correlation above. We derive an analytic model of the full monochromation process including these effects, and show this model is in excellent agreement with particle tracking simulations. We also derive the lower limit of achievable monochromation due to 2), which reveals the quantitative impact of 1) in realistic electron optical layouts. Finally, we evaluate the sensitivity of the monochromation to amplitude and phase jitter of the THz, and find that practical implementations of this technique are unlikely to be limited by jitter.

\begin{figure*}
    \centering
    \includegraphics[width=\linewidth]{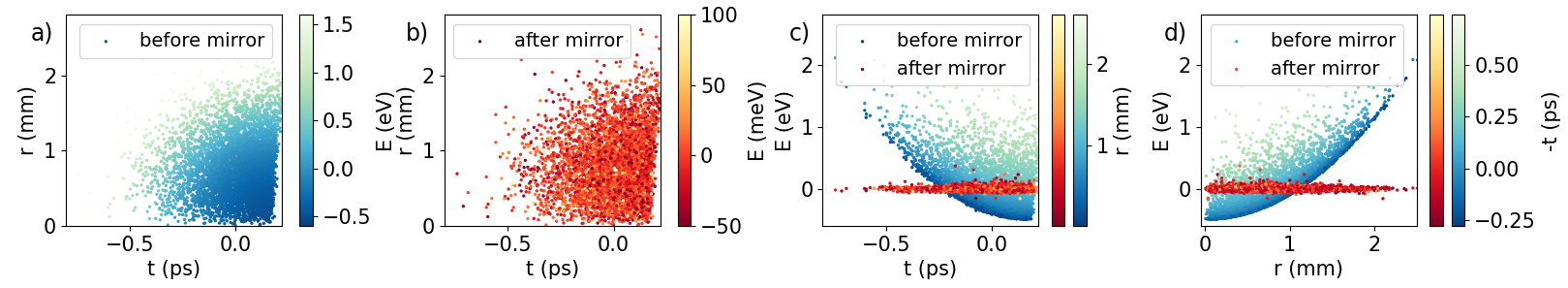}
    \caption{GPT particle distributions before and after the THz monochromator with two pulses at 0.3 THz. For the simulation shown here $\sigma_{x0} = 1$ $\mu$m and $\sigma_{t0} = 8.6$ fs.  $\boldsymbol{a)}$ Real space radius vs time of arrival colored by energy deviation (from the mean) before the mirror, showing the correlations that develop during transport. $\boldsymbol{b)}$ Real space energy deviation after monochromation. $\boldsymbol{c)}$ Particle energy vs time of arrival colored by radius. The blue-colored distribution is taken from before the mirror monochromator and the red-colored is from after. $\boldsymbol{d)}$ Particle energy vs radius colored by time of arrival.}
    \label{fig:dists}
\end{figure*}

The general layout we assume is of an electron beam emerging from a flat photocathode, in the single electron per pulse limit and so Coulomb interactions are avoided. Also in this limit the energy spread is determined by the source and growth in energy spread due to space charge effects is negligible. Electrons are assumed to be accelerated by a constant voltage electron gun, and focused with a round lens downstream. Our THz interaction is tilted mirror scheme, similar to the layout described in \cite{Ehberger2018-tilted}. The layout of the electron source, focusing lens, and interaction point is shown in Fig.~\ref{fig:schematics}a), the shape of a THz pulse is shown in, Fig.~\ref{fig:schematics}b), and cartoon of the mirror geometry is shown in Fig.~\ref{fig:schematics}c). While our analytic formalism does not depend on the beam energy, in our simulations we choose a value of 60~keV kinetic energy. The gun geometry is a flat cathode and anode electrode, with a cathode-anode gap of 10 mm and launch field of 6 MV/m, and an anode hole diameter of 3~mm, which sets its (de)focusing power. In Fig.~\ref{fig:dists} we show an example of particle tracking results, which displays the particle distribution in energy, space, and time for this simulation, and illustrates the overall effect of the monochromator. Details of these simulations will be presented in Sec.~\ref{sec:sims}.

\section{Correlation between energy, position, and arrival time}

We begin by deriving how energy becomes correlated with transverse radius and arrival time in an arbitrary linear optical system. We first consider the ideal limit that all particles are emitted instantaneously from the source at a single point, in which case the energy of a particle is perfectly correlated with time and space. We assume the beam has a finite spread in momentum such that it is a gaussian with standard deviation $\sigma_p$ in both transverse coordinates, and a forward half-gaussian in the longitudinal direction. If particles were emitted in all directions, the longitudinal momentum standard deviation would also be $\sigma_p$; by limiting to the forward half-space the standard deviation becomes $\sigma_p/\sqrt{2}$. At some longitudinal distance $\Delta z$ down the beamline a particle evolves to have transverse position $(x, y)$ and time of arrival $t$ (relative to the bunch centroid). Our model is that of a dc electron gun followed by transverse focusing optics. In the approximation that the transport is linear, that there is no coupling between the transverse planes, and that the applied fields are cylindrically symmetric, the final space-time coordinates are related to their initial momentum by
\begin{align} \label{xyt_transfer}
         &x = M_{12}p_{x0}, \quad y = M_{12}p_{y0}, \quad t = M_{56}p_{z0},
\end{align}
where $M_{ij}$ are the components of the $6\times6$ transport matrix defined in terms of the particle momentum, rather than the (more traditionally used) angle. This distinction is important at the particle source where the momentum is rapidly changing. Defining $E_{gun}$ as the amount of kinetic energy an electron gains inside the dc gun, we now express the final kinetic energy $E=E_{gun}+\frac{1}{2m}(p_{x0}^2+p_{y0}^2+p_{z0}^2)$ in terms the final positions:

\begin{align}
    E_{ideal}(r, t) = E_{gun} + \frac{r^2}{2m M_{12}^2} + \frac{t^2}{2m M_{56}^2} \label{Eideal}
\end{align}
where $r = \sqrt{x^2+y^2}$ is the transverse radius, and the subscript on $E_{ideal}$ labels it as the ideal, zero size emitter case. It is apparent that the energy distribution is a three dimensional paraboloid centered at $x,y,t=0$. Eq.~\ref{Eideal} is valid even for relativistic gun energies, provided that the electron emission energy is much smaller than the rest mass. The dc gun provides a constant energy boost to every particle, and therefore does not change the energy-time dispersion from what it is at the point of emission. The energy spread and energy-time/energy-radius dispersion also does not change in propagation, provided that space-charge effects are negligible. 

 The actual energy of a particle has some deviation from this paraboloid because of the finite emittance of the beam. In other words, when the rms emission size $\sigma_{x0}$ and rms laser pulse duration $\sigma_{t0}$ are nonzero there is a contribution to the energy distribution that is uncorrelated with space or time, and hence not taken into account by Eq.~\ref{Eideal}. We will show that this uncorrelated component is the absolute limit of how much the beam can be monochromated, and can in principle be made small in a practical electron optical system. Assuming this, significant monochromation can be achieved by applying a complementary parabolic energy gain in space and time. The associated applied force need only be in the longitudinal direction, as the large majority of the energy spread is due to longitudinal momentum spread after acceleration. The integral of this force over the time during which it acts on the electron beam must vary parabolically in transverse radius and time of arrival for ideal monochromation.

\section{Minimum achievable energy spread}

The uncorrelated energy spread sets a lower limit on the final $\sigma_{E}$ that can be achieved by removing energy correlations with a time- and space-dependent force, regardless of the details of the force itself. To compute this, we first assume we can generate a tunable, parabolic energy change $\Delta E = -ar^2-bt^2$, on each particle according to its time of arrival $t$ and transverse radius $r$, where $a$ and $b$ are tunable parameters constant across the bunch. After adding this energy change to a particle of energy $E$, the residual energy deviation, 
\begin{equation}
    E_{res} = E - a(x^2+y^2) - b t^2,
    \label{Eres}
\end{equation}
can be used to compute the rms final energy spread  $\sigma_{E,res}^2 = \langle E_{res}^2 \rangle - \langle E_{res} \rangle^2$, by averaging over the initial particle distribution. As above, the particle coordinates are  determined by their initial  momentum, but now also their initial position and emission time:
\begin{align} \label{xyt_transfer}
         &x = M_{11}x_0+M_{12}p_{x0}, \nonumber \\
         &y = M_{11}y_0+M_{12}p_{y0}, \nonumber \\
         &t = t_0+M_{56}p_{z0} .
\end{align}
The averages required to compute $\sigma_{E,res}$ are evaluated over the momentum and spatial/temporal distributions. The initial momentum distribution is assumed to be gaussian with variance $\sigma_{p0}^2=\frac{2}{3}m \sigma_{E0}$. After minimizing $\sigma_{E,res}$ with respect to the parameters $a$ and $b$ it becomes,
\begin{align}
    \sigma_{E,res}^2
    = \frac{1}{3} \sigma_{E0}^2 \Biggl(1-\frac{1}{(1+\alpha_t)^2} \Biggr) 
    + \frac{2}{3} \sigma_{E0}^2 \Biggl(1-\frac{1}{(1+\alpha_x)^2} \Biggr)
    \label{sigE_res}
\end{align}
where,
\begin{equation}
    \alpha_t = \sqrt{\frac{3}{2}} \frac{\sigma_{t0}^2}{m M_{56}^2 \sigma_{E0}}, \quad \alpha_x = \sqrt{\frac{3}{2}} \frac{M_{11}^2 \sigma_{x0}^2}{m M_{12}^2 \sigma_{E0}}.
\end{equation}
The geometric factors of $\sqrt{\frac{3}{2}}$ arise from integrating gaussian functions. We have assumed here that the beamline is cylindrically symmetric and $x$ and $y$ are uncoupled; the more general form of $\sigma_{E,res}$ is derived in Appendix A.

The first term of Eqn.~\ref{sigE_res} accounts for energy spread that arises because of the finite initial temporal length of the beam, while the second term accounts for that which is due to the finite size on the cathode. The matrix elements $M_{12}$ and $M_{56}$ represent the strength of the energy correlations, which means we can enhance the ability to monochromate the beam by choosing beamline parameters to maximize $M_{12}$ and $M_{56}$. But these choices often involve to tradeoffs with other preferable beam qualities. For example, decreasing the gun field for a given voltage will lead to a larger $M_{56}$, but at the same time increases the bunch length, sacrificing time resolution in an ultrafast experiment. And so the details will vary on an application-specific basis. It is also important to note that the percent reduction in energy spread is dependent on $\sigma_{E0}$, with $\sigma_{E,res}^2 / \sigma_{E0}^2 = 1$ as $\sigma_{E0} \rightarrow 0$ and $\sigma_{E,res}^2 / \sigma_{E0}^2 \sim \frac{1}{\sigma_{E0}}$ as $\sigma_{E0} \rightarrow \infty$.

The coefficients $a$ and $b$ from Eqn.~\ref{Eres} represent the best-fit paraboloid for a given energy distribution:
\begin{align}
    & a = \frac{2m M_{12}^2 \sigma_{E0}^2}{ (2mM_{12}^2 \sigma_{E0} + 3 M_{11}^2 \sigma_{x0}^2)^2 }, \label{a} \\
    & b = \frac{2m M_{56}^2 \sigma_{E0}^2}{ (2mM_{56}^2 \sigma_{E0} + 3 \sigma_{t0}^2)^2 } \, . \label{b}
\end{align}
In the zero-emittance limit where the energy is perfectly correlated, Eqns.~\ref{a} and \ref{b} reduce to the coefficients of the paraboloid in Eqn.~\ref{Eideal}. As the initial bunch length and size on the cathode become large, $a$ and $b$ become small, indicating a shallow paraboloid and weak energy correlations. In the next section we will consider how a THz pulse can generate an energy change that best approximates this two-dimensional paraboloid.

\begin{figure}
    \centering
    \includegraphics[width=\linewidth]{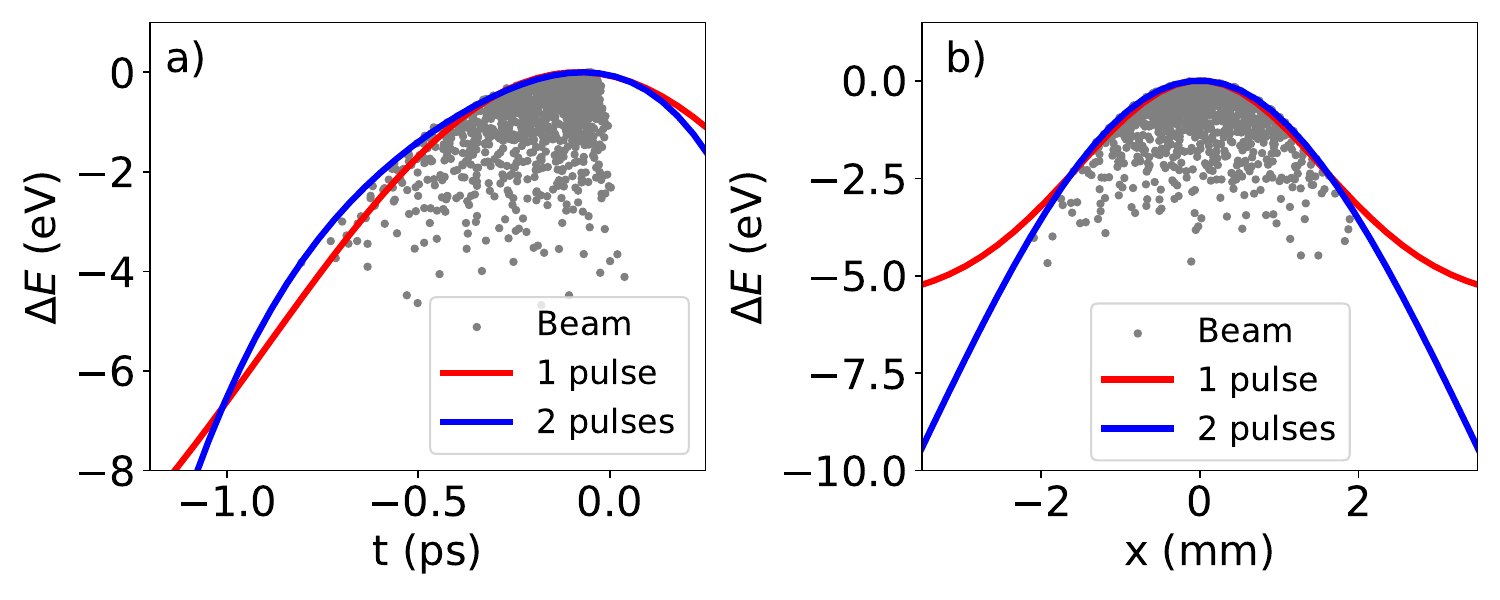}
    \includegraphics[width=\linewidth]{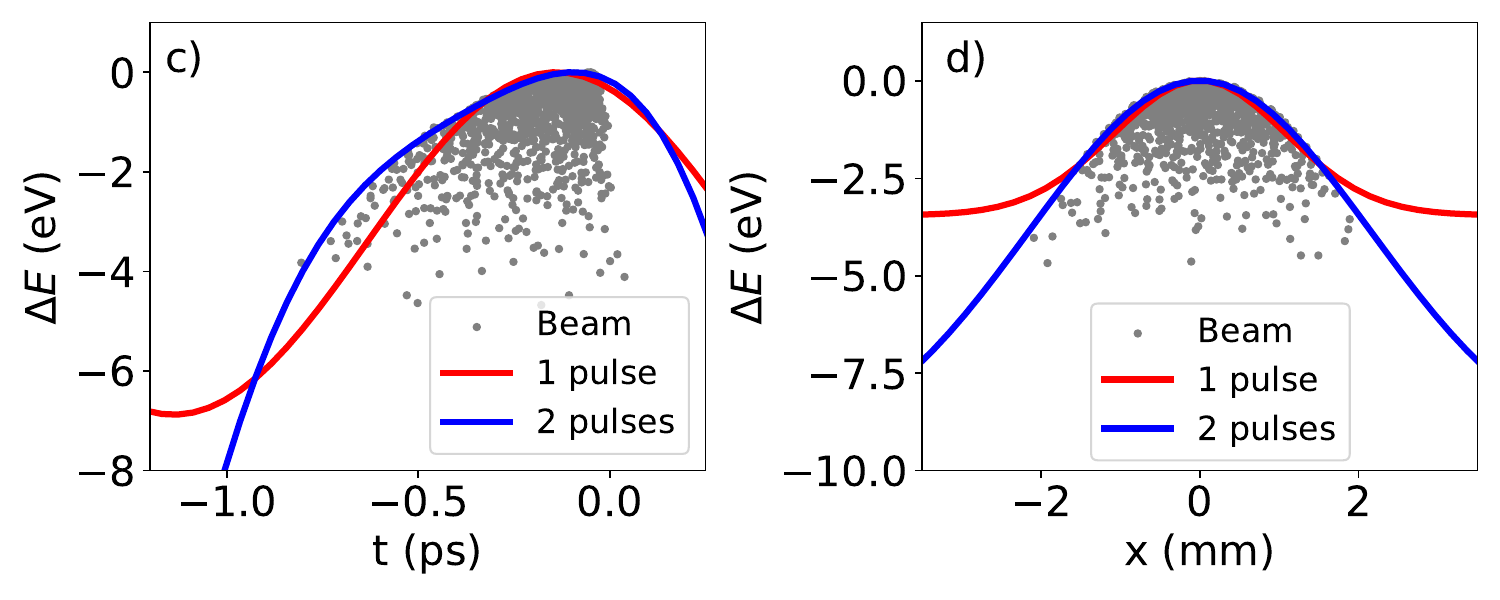}
    \includegraphics[width=\linewidth]{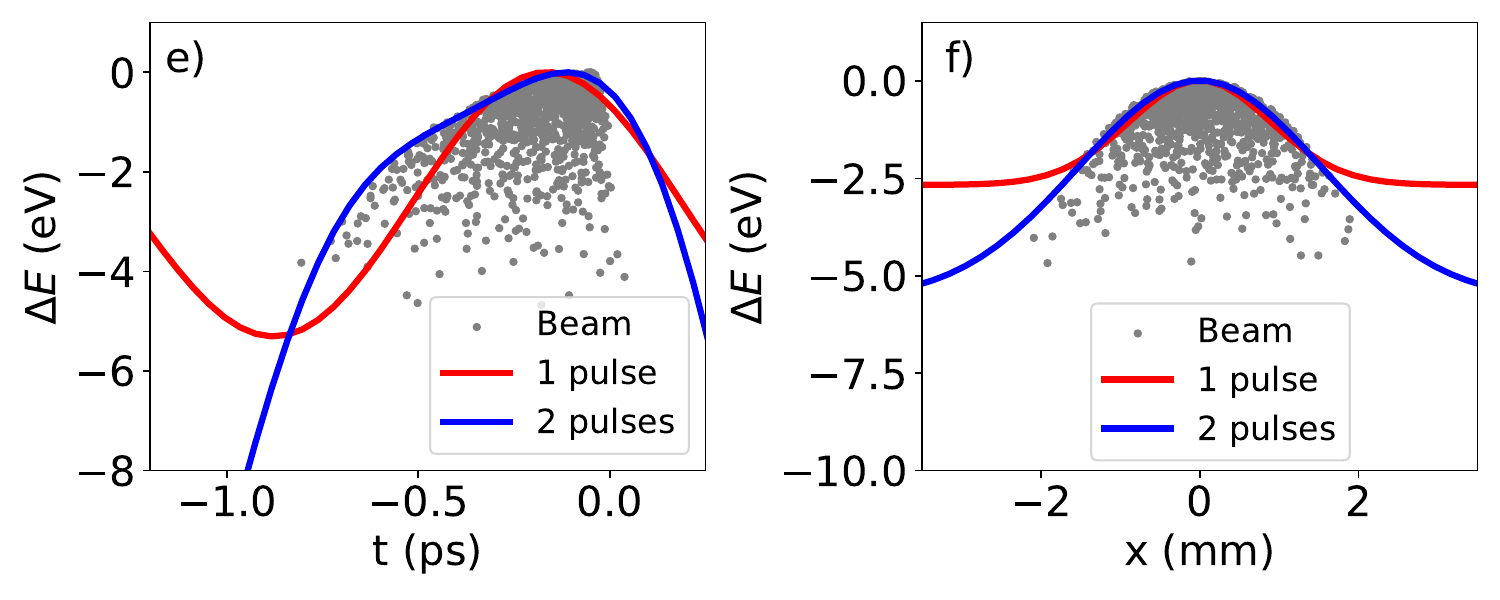}
    \caption{Energy change from 1 pulse (red) and two pulses (blue) for cross sections in time offset and radius for central frequencies at $\boldsymbol{a,b)}$ 0.3 THz, $\boldsymbol{c,d)}$ 0.5 THz, and $\boldsymbol{e,f)}$ 0.7 THz. The gray dots show the distribution from GPT simulations, with the negative energy plotted for better comparison to the energy change parabola. The cross sections in time are taken at $(x,y)=0$, and the cross sections in $x$ are taken at $(y,t)=0$. For every simulation shown here $\sigma_{x0} = 1$ $\mu$m and $\sigma_{t0} = 8.6$ fs. Here time of arrival is measured relative to a particle emitted from the cathode with zero longitudinal momentum, hence the sharp cutoff at t=0.}
    
    \label{fig:pulse_parabolas}
\end{figure}


\section{Analytic model of electron beam and terahertz interaction}

The analytic expression for the energy change induced by interaction with a THz pulse at a mirror barrier is derived in detail in Appendix \ref{Appendix_derivation} following a similar process to that described in reference \cite{Ehberger2018-tilted}, and here we will quote the results. The THz pulse is modeled as a gaussian wave packet where $E_0$ is the amplitude of the electric field at the origin, $w_0$ is the size of the beam waist, $\omega_0$ is the angular frequency, $\sigma_t$ is the rms pulse duration, and $\phi_0$ is an overall phase offset. The mirror is assumed to be perfectly reflective, and the electron trajectories are treated as straight line paths through the THz field. The total electric field is the superposition of the incident and reflected THz pulses, each at an angle $\theta_{T}$ relative to the mirror normal. The electron beam passes through the mirror at angle $\theta_B$ relative to the mirror normal, with velocity $\beta c$, and gains momentum equal to the integral of the force along its trajectory:
\begin{equation} \label{p_integral}
    \Delta \vec{p}=\int_{0}^{\infty} { Re \Bigl[ \vec{F}_\text{total} \left( x(t), y(t), z(t), t \right) \Bigr] dt}.
\end{equation}
Within the approximation of a narrow spectrum ($\omega_0 \sigma_t >> 1$), the paraxial limit ($ w_0 \omega_0 /c \gg 1$), and small enough forces to not perturb the electron trajectories, the resulting components of the momentum transfer are,
\begin{align} \label{dpz}
    \Delta p_z = &\frac{e E_0}{\omega_0} \times \exp{\left[{\frac{-1}{w_0^2} \left( y^2 + \left(x \frac{\cos{\theta_T}}{\cos{\theta_B}} \right)^2 \right)} \right] } \nonumber \\
    & \times \frac{2 \cos{\theta_B} (\beta \sin{\theta_B}-\sin{\theta_T}) }{\bigl(\beta \cos{(\theta_B-\theta_T)} - 1 \bigr) \bigl(\beta \cos{(\theta_B+\theta_T)} + 1 \bigr) } \nonumber \\
    &\times  \sin \left[ \phi_0 - \omega_0 \left(t - \frac{x ( \beta \sin{\theta_T} - \sin{\theta_B} )}{\beta c \cos{\theta_B}} \right) \right]
\end{align}

\begin{align} \label{dpx}
    \Delta p_x = &\frac{e E_0}{\omega_0} \times \exp{\left[{\frac{-1}{w_0^2} \left( y^2 + \left(x \frac{\cos{\theta_T}}{\cos{\theta_B}} \right)^2 \right)} \right] } \nonumber \\
    & \times \frac{2 (\beta \sin{\theta_T}-\sin{\theta_B})(\beta \sin{\theta_B}-\sin{\theta_T}) }{ \left(\beta \cos{(\theta_B-\theta_T)} - 1 \right) \left(\beta \cos{(\theta_B+\theta_T)} + 1 \right) } \nonumber \\
    & \times \sin \left[ \phi_0 - \omega_0 \left(t - \frac{x ( \beta \sin \theta_T - \sin{\theta_B} )}{\beta c \cos{\theta_B}} \right) \right].
\end{align}

Here, the coordinate system is the beam's coordinates, so $\Delta p_z$ is a longitudinal momentum kick, $\Delta p_x$ is transverse, and the electron's coordinates $(x, y, t)$ are relative to an electron that is incident on the center of the mirror just as the center of the THz pulse arrives. As shown in  Appendix \ref{Appendix_derivation}, these approximate analytic expressions agree remarkably well with the momentum change found by numerically integrating Eqn.~\ref{p_integral}, even for few cycle pulse durations. An important condition occurs when the electron beam's angle of incidence is chosen to satisfy
\begin{equation} \label{angles}
\beta \sin{\theta_T} = \sin{\theta_B}.
\end{equation}
In this case, two simplifications happen. First, the longitudinal momentum change $\Delta p_z$ becomes an even function in $x_0$, removing that asymmetry. Second, the transverse momentum change $\Delta p_x$ goes to zero, allowing us to write the total energy change as $\Delta E = \beta c \Delta p_z$. We assume this condition is met for the remainder of this work.

To make more apparent how $\Delta E$ from the THz pulse counteracts parabolic energy correlations, we Taylor expand $\Delta E = \beta c \Delta p_z$ about $(x, y, t) = (0, 0, 0)$. 
\begin{align} \label{dE_symm}
    \Delta E & \approx A \left( 1 - \frac{x^2}{w_0^2} \frac{\cos^2{\theta_T}}{1-\beta^2 \sin^2{\theta_T}} - \frac{y^2}{w_0^2} - \frac{t^2 \omega_0^2}{2} \right) \nonumber \\
    & \approx A \left( 1 - \frac{r^2}{w_0^2} - \frac{t^2 \omega_0^2}{2} \right) \\
    & \text{for} \, \theta_T \ll 1 \nonumber \\
    \text{with} &\quad A = \frac{e E_0 2 \beta c \sin \theta_T}{\omega_0 \sqrt{1 - \beta^2 \sin^2 \theta_T}} \nonumber
\end{align}

For small THz incident angles, the energy change evidently takes the form of a negative parabola in $r$ and $t$. Importantly, the coefficients in front of $r^2$ and $t^2$ can be tuned independently of one other using the two relevant degrees of freedom for a laser, amplitude and focal size. The curvature in time is determined by adjusting the amplitude $E_0$, leaving the focal size $w_0$ to match the desired curvature in $r$. Therefore, to second order, a single THz pulse can perfectly correct energy correlations in an electron beam. This idealized case is precisely the limit described by Eqn.~\ref{sigE_res}. However, as the temporal length of the electron bunch approaches the period of one half-cycle of a THz pulse, the parabolic approximation of the momentum change becomes less appropriate and contributions from higher order terms are significant. Fig. \ref{fig:pulse_parabolas} shows the energy change as a function of arrival time and radius for three example THz frequencies. The parameters $E_0$ and $w_0$ were determined based on a numerical minimization of the final energy spread for a simulated distribution of particles (which will be described in detail in the next section). The gray distribution shows the negative energy of the particles, and represents the desired energy change. In other words, maximal monochromation occurs when the curvature of $\Delta E$ matches the curvature of the particle energies. This happens at low THz frequency, where also the energy change from one pulse and two pulses are not significantly different. In the time dimension, we can see higher order terms becoming problematic as the frequency grows larger and the curvatures begin to differ. 

Monochromation in the radial dimension is not limited by higher order terms in the same way (as there is no hard limit on the transverse size of a laser), and the curvature can be matched well by one pulse regardless of frequency with a strategic choice of spot size. With frequency fixed, amplitude is the only degree of freedom that influences the temporal dimension. But amplitude cannot be used in isolation to optimize temporal curvature due to its simultaneous effect on the radial distribution. For example, we can see visually in Fig.~\ref{fig:pulse_parabolas}e with one pulse at 0.7 THz the temporal curvature would be better matched with a smaller amplitude, but would then result in insufficient field strength at the radial edges of the beam (Fig.~\ref{fig:pulse_parabolas}f). The addition of a second pulse provides a second degree of freedom in time, allowing the amplitude to more closely match the temporal curvature without compromising the effect in the radial dimension.

This result is notably different from the previous proposal using rf cavities in which the $\Delta E$ curvatures in $r$ and $t$ are opposite in sign, necessitating the use of two cavities with opposite amplitudes. The convenience of the THz mirror is that it is capable of reducing energy spread in both dimensions simultaneously with a single pulse. In the case that higher order terms are significant, the addition of a second pulse reflected from the same mirror adds minimal complication to a design, in contrast to the rf cavity approach which requires at least two interaction points along the beamline. 


\section{Particle simulations} \label{sec:sims}

In this section we describe the particle tracking simulations used to model the THz monochromator, which were performed in General Particle Tracer (GPT) \cite{gpt}. The GPT simulation consisted of a simple analytic defocusing DC gun at 60 kV, a collimating solenoid, and the THz mirror. The initial distribution at the cathode was chosen to resemble that of a realistic state of the art electron source, with a radial size $\sigma_{x0} = 1$ $\mu$m and a duration of 30 fs ($\sigma_{t0} = 8.6$ fs). The distributions seen in Fig.~\ref{fig:dists} are taken from just before the mirror, and show a distinctly parabolic trend in space and time, just as predicted by Eqn.~\ref{Eideal}. 

\begin{figure}[h!]
    \centering
    \includegraphics[width=0.85\linewidth]{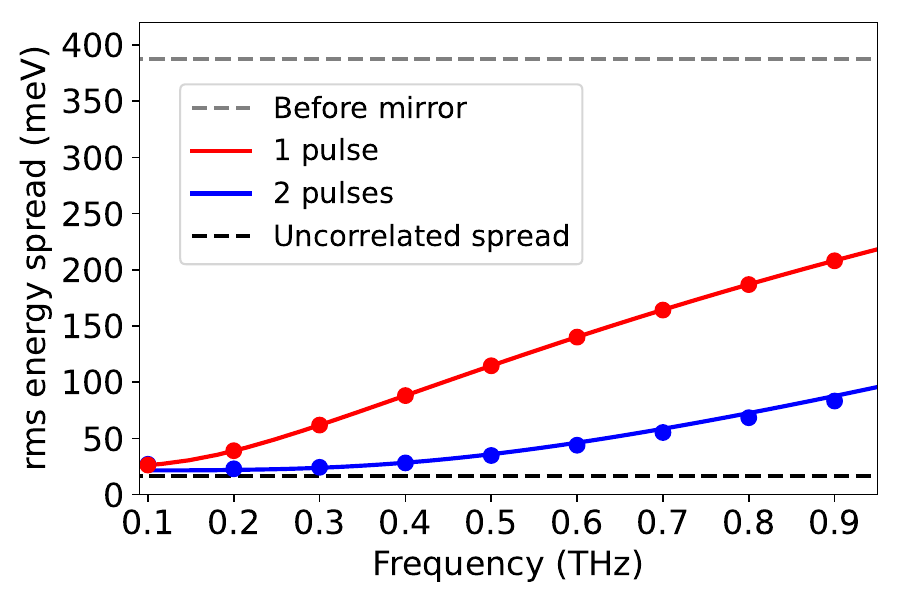}
    \caption{Final energy spread after the THz interaction as a function of THz frequency. The solid lines are found by minimizing $\sigma_E$ with respect to the parameters $E_0$ and $w_0$ in Eqn.~\ref{dpz}. The points correspond to the result of a GPT simulation using the optimal parameters from the analytic optimization at that frequency. For every simulation shown here $\sigma_{x0} = 1$ $\mu$m and $\sigma_{t0} = 8.6$ fs, and the distribution just before the mirror is shown in Fig.~\ref{fig:dists}. THz pulse parameters for each point are described in Appendix~\ref{Appendix_parameters}.}
    \label{fig:best_sigE}
\end{figure}

We modeled the effect of the THz mirror on a distribution of particles in two separate ways; first by calculating the energy change on each individual particle using the analytic relations described in section IV, and second by including the THz interaction as a custom GPT element with the true electric and magnetic fields of the THz pulses as described in Appendix~\ref{Appendix_derivation}. We performed a multi-variable numerical minimization on the analytic function to determine the pulse parameters that yielded the smallest energy spread, then used those same parameters for the THz interaction step in GPT. We chose to model frequencies between 0.1 THz and 1 THz, a range in which the half-wavelength is longer than the temporal duration of a typical pulsed electron beam, and in which THz radiation can be produced by optical rectification. The number of cycles in the pulse was held constant between frequencies and the phase was fixed at 90 degrees for every simulation. The THz pulse and electron beam have angles of incidence $\theta_T$ of 20 degrees and $\theta_B$ of 5.8 degrees relative to the mirror normal, which satisfy the condition described by Eq.~\ref{angles}. For two THz pulses the amplitude, width, and time delay are allowed to vary separately. Fig.~\ref{fig:best_sigE} shows the best final energy spread for one pulse or two pulses as a function of frequency. The solid lines are the final energy spread according to the analytic model after minimizing with respect to the THz pulse amplitude, width, and time delay. The points represent GPT simulations of the mirror element with the same parameters from the analytic optimization at that frequency. The results from each method are almost identical. At small frequencies, the energy spread approaches the limit set by the uncorrelated energy spread found in Eqn.~\ref{sigE_res}.

For one pulse there exists a straightforward optimal combination of parameters at each frequency, which are shown in Fig.~\ref{fig:1pulseParams} in Appendix~\ref{Appendix_parameters}. For two pulses the optimization is under-constrained, and we find that the final energy spread is not strongly dependent on the pulse energy, but that pulse energy and $\sigma_E$ are competing variables. The final energy spread can always be made smaller by increasing the energy of the pulses, but asymptotes to its minimum value relatively quickly (see Fig.~\ref{fig:2pulseFronts} of Appendix~\ref{Appendix_parameters}). Hence, we anticipate that THz energies on the order of tens of nanojoules would be adequate. Details on the pulse energy and the optimization parameters are described in Appendix C.

\section{Stability}

\begin{figure}[h!]
    \centering
    \includegraphics[width=0.65\linewidth]{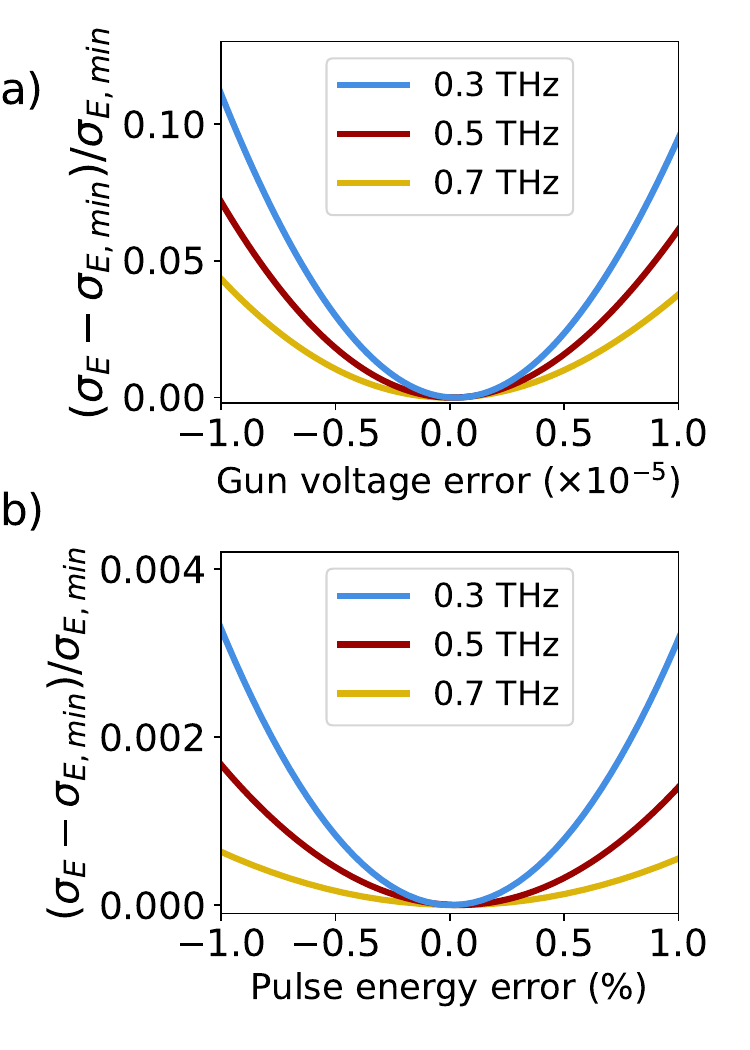}
    \caption{Energy spread after fluctuations in $\boldsymbol{a)}$ gun voltage and $\boldsymbol{b)}$ terahertz pulse energy for two pulses at 0.3, 0.5, and 0.7 THz. }
    \label{fig:stability}
\end{figure}

Fluctuations in the THz field amplitude and the electron source voltage can both hinder the performance of the monochromator. Although in our model the energy of the electrons is irrelevant, error in energy causes variations in the arrival time of the pulses; in the small region of variation that we are concerned about, the relationship between energy error and arrival time is linear. We re-evaluated the analytic model while varying these parameters for a two pulse monochromator with center frequencies 0.3, 0.5, and 0.7 THz, and found the resulting energy spread to be only marginally affected (Fig.~\ref{fig:stability}). High voltage power supplies for electron microscopy can have relative stability better than $10^{-5}$ which at 60 keV in our case corresponds to a time delay of 15 fs. We assume a representative THz energy fluctuation of  $\sim \pm 1 \%$. At this level of fluctuation, gun voltage error would increase the final energy spread by less than 15\%, and pulse energy error by less than 0.5\%. The relative increase in energy spread is larger for lower frequencies because the center (no jitter) energy spread is smaller. Note that we assume the time of arrival fluctuation in the THz radiation is much smaller than that associated with the electron arrival time jitter due to voltage fluctuations.

\section{Reflection from a mesh mirror}

Until this point we have assumed the mirror to be a perfectly reflective conductor, and for the most part ignored the mechanism by which the electrons are able to pass through the mirror. We now relax this assumption to allow for transmission of a real electron beam, choosing a mesh grid as a model for a mirror that allows significant electron transmission, while having near zero THz transmission. For higher energy beams it may be more suitable to use a thin metal membrane for a mirror, similar to experiments in Ref. \cite{Morimoto2018}, \cite{Murooka2011}, and \cite{Ehberger2019}. However the particle loss through a membrane grows as the beam energy becomes smaller. We choose to model a mesh grid, for which the percent transmission can be large even for low energy beams. 

\begin{figure}
    \centering
    \includegraphics[width=0.5\textwidth]{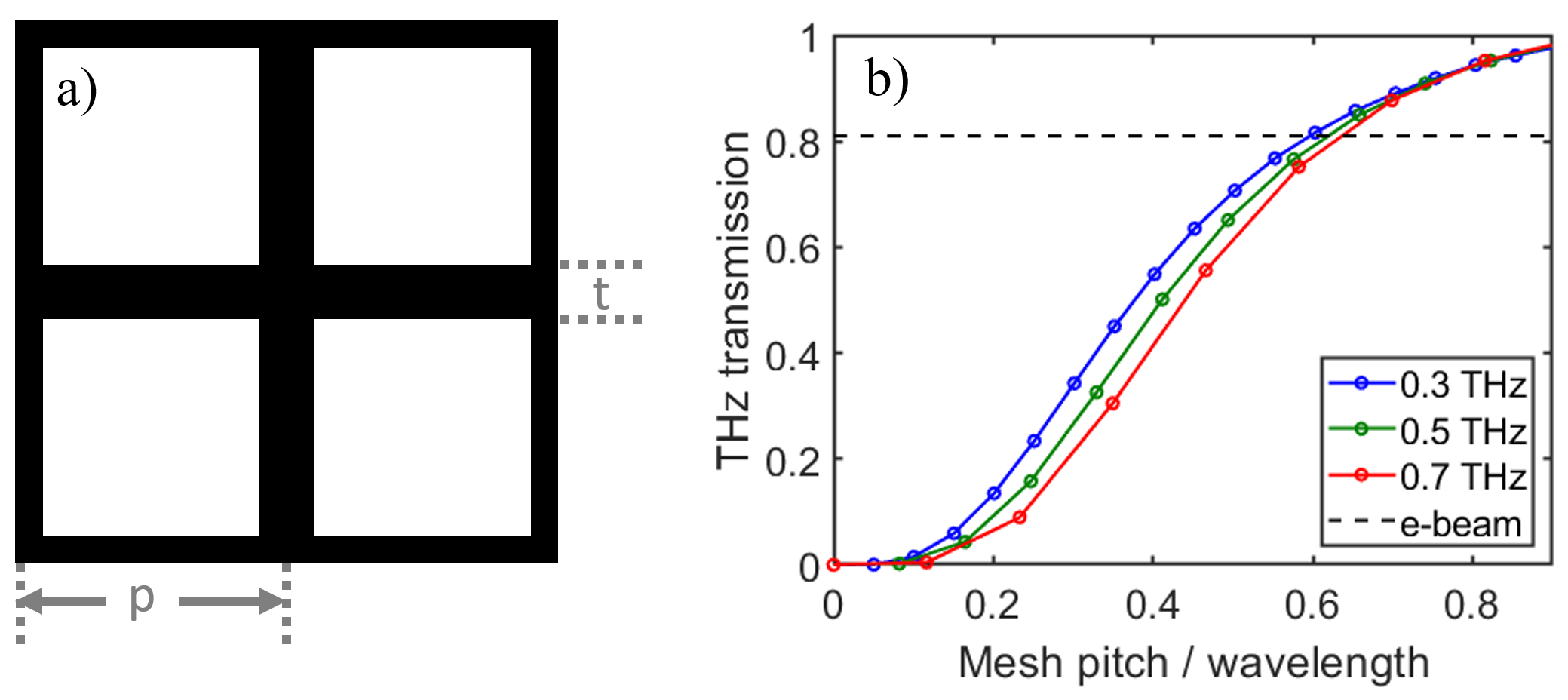}
    \caption{a) The geometry of the simulated wire mesh. The wires have thickness $t$ and pitch $p$, and the ratio $t/p = 0.1$ was held constant. b) Transmission of a normal incident THz wave at three different frequencies as a function of $p/\lambda$. Also shown is the transmission of the electron beam, assumed to be the fraction of empty space in the mesh, $T_e = (1-t/p)^2$.}
    \label{fig:mesh_transmission}
\end{figure}

We use Ansys Lumerical FDTD \cite{lumericalFDTD} to model the THz reflection. We calculate the transmission as a function of the mesh pitch $p$, but keep the ratio of the bar thickness $t$ to the pitch a constant, $t/p = 0.1$ to ensure that the transmission of the electron beam is a constant value $T = (1-0.1)^2 = 0.81$. The longitudinal thickness of the mesh is also held at a constant 20 micron, chosen to match commercially available TEM grids, and we approximate the mesh material as a perfect electrical conductor. The geometry is shown in Fig.~\ref{fig:mesh_transmission}a. We look only at the case of normal incidence of the THz pulse, which is likely a worst case, since the apparent pitch will only decrease for non-normal incidence. We measure the transmission of a broad spectrum THz wavepacket, and sample the results at chosen frequencies. In Fig.~\ref{fig:mesh_transmission}b we plot the results for three example THz frequencies. As expected, when the mesh pitch is well below the THz wavelength, the mesh acts almost like a perfectly reflecting barrier. In this region of negligible THz transmission we also find that the momentum kick (calculated numerically from the Lumerical simulation fields) is nearly identical to that from a perfectly reflecting mirror, indicating that there are no stray fields inside the mesh holes that would distort the electron beam's trajectory. Having a mesh pitch that is well below the wavelength is ideal not only for maximum reflection of the THz pulse, but also to avoid resonances that occur when the pitch is of comparable size to the wavelength.


\section{Conclusion}

We have shown that laser-derived THz radiation can be used to correct energy spread in an electron beam, by exploiting the natural parabolic energy-space correlations that arise from a source with finite momentum spread. We predict that an rms energy spread reduction of more than a factor of ten is possible for a beam with initial energy spread of hundreds of meV, particularly when using two THz pulses, yielding rms energy spreads of a few 10s of meV. We determine analytically and in simulation the conditions necessary to correct energy spread correlated with both time and transverse position, and show excellent agreement between the two. We show that the effects of jitter are unlikely to dominate the final energy spread in practice. Finally, we show that by using commercially available wire meshes, the amount of current loss in the electron beam can be limited to a few 10s of percent, thereby significantly outperforming prism-based monochromators in this regard. 

\section{Acknowledgments}
This work was supported by DOE awards DE-SC0024907, DE-SC0020144, and the National Science Foundation Grant No. PHY-1549132, the Center for Bright Beams.

\clearpage

\appendix

\section{Residual energy from a generalized transfer matrix}

The general transfer equations for $x$, $y$, and $t$, allowing for coupling of the transverse coordinates and not requiring cylindrical symmetry is,
\begin{align} 
    & x = M_{11}x_0 + M_{12}p_{x0} + M_{13}y_0 + M_{14}p_{y0} \label{xgeneral} \\
    & y = M_{11}x_0 + M_{12}p_{x0} + M_{13}y_0 + M_{14}p_{y0} \label{ygeneral} \\
    & t = t_0 + M_{56}p_{z0}.
\end{align}
We solve Equations \ref{xgeneral} and \ref{ygeneral} as a system for $p_{x0}$ and $p_{y0}$, and follow the same procedure as in section III. We find the residual energy spread to be,

\begin{multline}
    \sigma_{E0}^2 = \frac{1}{3} \sigma_{E0}^2 \Biggl( 1-\frac{1}{(1+\alpha_t)^2} \Biggr) \\
    + \frac{2}{3} \sigma_{E0}^2 \Biggl( 1-\frac{(M_{12}^2+M_{14}^2+M_{21}^2+M_{34}^2)^2}{A_{x1}+A_{x2}+A_{x3}} \Biggr)
\end{multline}

where,

\begin{align}
    & A_{x1} = 2 \bigl( (M_{12}^2+M_{14})^2 \\ \nonumber
    & \quad \quad \quad + 2(M_{12}M_{32}+M_{14}M_{34})^2 \\ \nonumber 
    & \quad \quad \quad + (M_{32}^2+M_{34})^2 \bigr) \\
    & A_{x2} = 2 \sqrt{5} \bigl( (M_{11} M_{12} + M_{31} M_{32})^2 \\ \nonumber
    & \quad \quad \quad + (M_{12} M_{13} + M_{32} M_{33})^2 \\ \nonumber
    & \quad \quad \quad + (M_{11} M_{14} + M_{31} M_{34})^2 \\ \nonumber
    & \quad \quad \quad + (M_{13} M_{14} + M_{33} M_{34})^2 \bigr) \frac{\sigma_{x0}^2}{m \sigma_{E0}} \\
    & A_{x3} = 3 \bigl( (M_{11}^2+M_{13})^2 \\ \nonumber
    & \quad \quad \quad + 2(M_{11}M_{31} + M_{13}M_{33})^2 \\ \nonumber
    & \quad \quad \quad+ (M_{31}^2+M_{33})^2 \bigr) \frac{\sigma_{x0}^4}{m^2 \sigma_{E0}^2}
\end{align}

\section{Derivation of energy transfer} \label{Appendix_derivation}

In this section, we derive a formula for the momentum kick given to an electron beam by a THz pulse reflecting off of a mirror, with geometry depicted in figure \ref{fig:schematics}a.  The starting point of the derivation is the Gaussian beam solution to the paraxial wave equation. This is a solution for a single frequency $\omega_0$, and will need to be modified to allow for a broadband pulse. To define the electric field, we will temporarily use a coordinate system such that the field is propagating in the z direction and has a (constant) polarization in the x-y plane. The scalar amplitude of the field is then given by 
\begin{equation}
E(r,z,\omega) = \frac{w_0}{w(z)} e^{\frac{-r^2}{w(z)^2}} e^{-i \left(\frac{\omega z}{c} + \frac{\omega r^2}{2 c R(z)} + \tan^{-1} \left(\frac{z}{z_r} \right) \right)}
\end{equation}
where
\begin{equation}
w(z) = w_0 \sqrt{1 + \frac{z^2}{z_r^2}} , \, R(z) = \frac{z^2 + z_r^2}{z}, \, z_r = \frac{\pi w_0^2 \omega}{2 c}
\end{equation}
The physical electric field is understood to be the real part of the complex field,
\begin{equation}
E_{\text{physical}} = \mathrm{Re} \left[ E(r,z,\omega) e^{i \omega t} \right].
\end{equation}
The solution for a broadband pulse of light can be written as a superposition of many of these single frequency solutions using a Fourier transform,
\begin{equation}
E_\text{THz} = \frac{1}{2 \pi} \int d \omega E(r, z, \omega) e^{i \omega t} \int dt' e^{- i \omega t'} E_\text{pulse}(t') \label{Eq:efield_exact}
\end{equation}
where $E_\text{pulse}(t)$ is the desired time dependence of the pulse. Because $E(0, 0, \omega) = 1$, this will remain the exact time dependence at the origin, even after the Fourier transforms. For this derivation, we will use a Gaussian pulse in time, with angular frequency $\omega_0$, pulse length $\sigma_t$, and an overall phase $\phi_0$,
\begin{equation}
E_\text{pulse}(t, \sigma_t) = e^{-\frac{t^2}{4 \sigma_t^2}} e^{i (\omega_0 t + \phi_0)}.
\end{equation}
The total energy in this pulse can be evaluated analytically by integrating the Poynting vector $\vec{S} = (1/\mu_0) \vec{E} \times \vec{B}$ at z=0 over all r and t,
\begin{equation}
\begin{aligned}
U_\text{pulse} & = \frac{2 \pi}{\mu_0 c} \int_0^\infty r dr \, e^{-2 r^2 / w_0^2} \int_{-\infty}^\infty dt \, \mathrm{Re}\left[E_\text{pulse}(t)\right]^2\\
& = \frac{\pi^{3/2} w_0^2 \sigma_t \left(1 + e^{- 2 \sigma_t^2 \omega_0^2} \cos(2 \phi_0) \right)}{2 \sqrt{2} c \mu_0}. \label{pulseEnergy}
\end{aligned}
\end{equation}

Up until now, no approximations beyond the paraxial wave approximation have been made, and if desired, the form of the field in Eq.\ \ref{Eq:efield_exact} could be used directly for numerical solutions. To proceed analytically, we will need to make a few approximations to. First, if the spectrum of the pulse is sufficiently narrow, then we can expand the integrand around $\omega=\omega_0$. We expand the amplitude and phase of the integrand differently, such that the phase is goes to second order, while the amplitude is just evaluated at $\omega_0$,
\begin{equation}
\begin{aligned}
& E(r, z, \omega) \equiv A(r, z, \omega) e^{i \phi(r, z, \omega)} \\
& \approx A(r, z, \omega_0) e^{i (\phi(r, z, \omega_0) + \frac{z}{c} \alpha_1 (\omega - \omega_0) + \alpha_2 (\omega-\omega_0)^2)} \\
& = E(r, z, \omega_0) e^{i (\frac{z}{c} \alpha_1 (\omega - \omega_0) + \alpha_2 (\omega-\omega_0)^2)}
\end{aligned}
\end{equation}
with the following definitions of $\alpha_1$ and $\alpha_2$,
\begin{equation}
\begin{aligned}
\alpha_1 & = 1 + \frac{8 c^4 (r^2 - \pi w_0^2) z^2 - 2 c^2 \pi^2 w_0^4 (r^2 + \pi w_0^2) \omega_0^2}{(4 c^2 z^2 + \pi^2 w_0^4 \omega_0^2)^2} \\
\alpha_2 & = \frac{8 c^3 \pi^2 w_0^4 (\pi w_0^2 - 3 r^2) z^3 \omega_0 + 2 c \pi^4 w_0^8 (r^2 + \pi w_0^2) z \omega_0^3}{(4 c^2 z^2 + \pi^2 w_0^4 \omega_0^2)^3}.
\end{aligned}
\end{equation}
The inverse Fourier transform can now be performed analytically to produce
\begin{equation}
E_\text{THz} = \frac{\sigma_t}{\tilde{\sigma}_t} E(r, z, \omega_0) e^{i  \frac{\omega_0 z \alpha_1}{c}} E_\text{pulse}\left(t - \frac{z \alpha_1}{c} , \omega_0, \tilde{\sigma}_t \right) \label{Eq:EThz1}
\end{equation}
with $\tilde{\sigma}_t = \sqrt{i \alpha_2 + \sigma_t^2}$. This is the form of the force that we use in GPT simulations of the THz field. 

\begin{figure*}[h!]
    \centering
    \includegraphics[width=\textwidth]{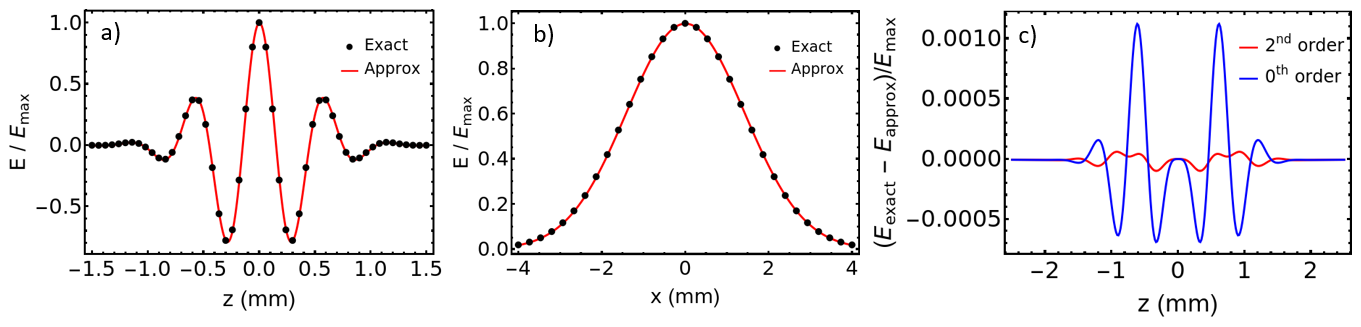}
    \caption{Electric field for an $f=0.5$ THz pulse at $t=0$ using either the exact formula in Eq.\ \ref{Eq:efield_exact} or the $2^\text{nd}$ or $0^\text{th}$ order approximations in Eqs. \ref{Eq:EThz1} and \ref{Eq:EThz2}, respectively. For these plots, $w_0 = 2$ mm, $\sigma_t = 1$ ps. (a) Field along the z-axis at $r=0$. (b) Field along the x-axis at $y=z=0$. (c) Difference between exact and approximate fields in (a) for both approximations. Well under $1\%$ error is achieved for either approximation.}
    \label{fig:field_comparison}
\end{figure*}

If we take the Taylor series of the phase only to 0$^{\text{th}}$ order such that $\alpha_1 \approx 1$, $\alpha_2 \approx 0$, the field is
\begin{equation}
E_\text{THz} = E(r, z, \omega_0) e^{i  \frac{\omega_0 z}{c}} E_\text{pulse}\left(t - \frac{z}{c} , \omega_0, \sigma_t\right). \label{Eq:EThz2}
\end{equation}
In Fig. \ref{fig:field_comparison}, the 2$^{\text{nd}}$ and 0$^{\text{th}}$ order approximations are compared to a numerical evaluation of the exact expression in Eq.\ \ref{Eq:efield_exact}. The analytic expressions are accurate to better than $1\%$  for either approximation, even for the case of a few cycle long pulse.

With this definition of propagating THz pulse fields, we can now calculate the fields of a THz pulse reflecting from a mirror. Shifting to the primed coordinate system in Fig.~\ref{fig:schematics}a where the mirror is now aligned along the $x'$ axis, and the THz is incident in the region $z' > 0$ at an angle $\theta_T$, we write the amplitude of the incident field in these new, rotated coordinates as,
\begin{equation}
E_\text{inc} = E_\text{THz}(x'\cos\theta_T - z'\sin\theta_T, x'\sin\theta_T + z'\cos\theta_T, t).
\end{equation}
The total field is written as the sum of the incident and reflected waves. We will restrict ourselves to the case of P polarization, so $E_{y'}=B_{x'}=B_{z'}=0$. The case of S polarization can be done similarly, but will not be considered here. In terms of $E_\text{inc}$, the total fields can be written as
\begin{equation}
\begin{aligned}
& z' \ge 0 : \\
& \quad E_{x'} = \cos\theta_T ( E_\text{inc}(x', z', t) - E_\text{inc}(x', -z', t)), \\
& \quad E_{z'} = -\sin\theta_T ( E_\text{inc}(x', z', t) + E_\text{inc}(x', -z', t)), \\
& \quad B_{y'} = \frac{1}{c} ( E_\text{inc}(x', z', t) + E_\text{inc}(x', -z', t)), \\
& z' < 0 : \\
& \quad E_{x'} = E_{z'} = B_{y'} = 0.
\end{aligned}
\end{equation}

Finally, we are ready to evaluate the forces felt by the electrons as they intersect with the THz pulse. We aim to evaluate the transverse and longitudinal momentum kicks on the beam in its own coordinate system, and we can relate the corresponding forces to the fields in the primed coordinate system by, 
\begin{equation}
\begin{aligned} 
F_{x} & = E_{x'} \cos\theta_B  - E_{z'} \sin\theta_B  - v_e B_{y'} \\
F_{z} & = E_{z'} \cos\theta_B  + E_{x'} \sin\theta_B  \\
\end{aligned}
\end{equation}
where $v_e = \beta c$ is the velocity of the electron. We approximate the trajectory of the electrons as a straight path through the THz field,
\begin{equation}
\begin{aligned}
x'_e & = \beta c\, (t - t_0) \sin\theta_B + x_0 \cos\theta_B\\
y'_e & = y_0 \\
z'_e & = \beta c\, (t - t_0) \cos\theta_B - x_0 \sin\theta_B.
\end{aligned}
\end{equation}

These trajectories are chosen so that in the beam coordinate system, $x(t) = x_0$ and $y(t) = y_0$ are constant, and $t_0$ is defined such that a particle with $x_0=0$ will cross the mirror at $t=t_0$. In general, the electron crosses the mirror at $t=t_0 + x_0 \tan\theta_B / \beta c$, so all integrals of the force will begin at this time, which we call the interaction time $t_\text{int}$.

\begin{figure*}[h!]
    \centering
    \includegraphics[width=\textwidth]{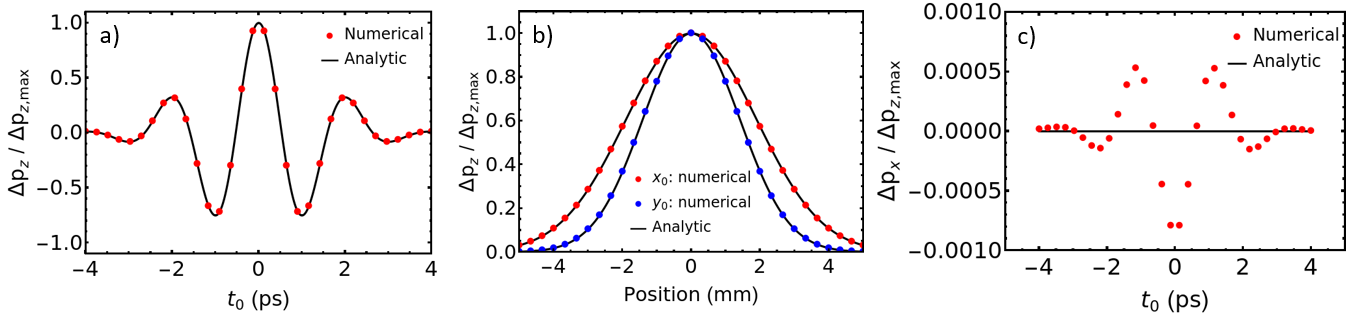}
    \caption{Comparison of the analytic expression for the momentum kick (lines) to a direct numerical integration of the forces (dots). For these plots, $f=0.5$ THz, $w_0 = 2$ mm, $\sigma_t = 1$ ps, $E_\text{beam} = 60$ KeV, $\theta_B$ is chosen to minimize the transverse kick, and $\theta_T = 45^o$ (a large angle to make the x-y asymmetry more obvious). (a) Temporal dependence of $\Delta p_z$ at $x_0=y_0=0$. (b) The spatial dependence of $\Delta p_z$ for both $x_0$ (red) and $y_0$ (blue). This x-y asymmetry is greatly reduced for $\theta_T \lesssim 20^o$. (c) Temporal dependence of $\Delta p_x$ at $x_0=y_0=0$, showing that the true transverse kick is not identically zero, but is still much smaller than the longitudinal kick.}
    \label{fig:momentum_comparison}
\end{figure*}

Finally, one last approximation must be made in order to be able to analytically evaluate the momentum kick. If the force is evaluated at $t=t_\text{int}$ and the paraxial approximation $ w_0 \omega_0 /c \gg 1$ is used to approximate the expression, it can be simplified to the following form,
\begin{equation}
F(x',z',t_\text{int}) \approx e^{-\frac{y_0^2 + x_0^2 \cos^2 \theta_T \sec^2 \theta_B}{w_0^2}} F_{\text{pw}}(x',z',t_\text{int}),
\end{equation}

where we have introduced the function $F_{\text{pw}}(x',z',t)$, which is the force in the plane-wave limit ($w_0 \to \infty$). Even though this is derived only for times near $t=t_\text{int}$, we will assume that this formula remains approximately valid for all times, or at least that most of the significant forces occur close to that time. This dramatically simplifies the momentum kick integrals, as all of the complexity from the focusing of the Gaussian beam has been buried in a single constant factor that only depends on the electron's spatial offset $(x_0, y_0)$. Note that while the fields are now infinite spatially, they still have a finite duration in time. With this plane-wave approximation, the forces can now be integrated from $t=t_\text{int}$ to $\infty$, producing the following expressions,

\begin{widetext}
\begin{equation}
\begin{aligned}
& \Delta p_z = G(x_0, y_0) H(x_0, t_0) \sqrt{\pi} \, \sigma_t \, e^{-\sigma_t^2 \omega_0^2}  \frac{\cos \theta_B \left( \beta \sin \theta_B - \sin \theta_T \right)}{\left(1 - \beta \cos(\theta_B - \theta_T)\right) \left(1 + \beta \cos(\theta_B + \theta_T)\right)},  \\
& \Delta p_x = G(x_0, y_0) H(x_0, t_0) \sqrt{\pi} \, \sigma_t \, e^{-\sigma_t^2 \omega_0^2} \frac{\left( \beta \sin \theta_T - \sin \theta_B \right) \left( \beta \sin \theta_B - \sin \theta_T \right)}{\left(1 - \beta \cos(\theta_B - \theta_T)\right) \left(1 + \beta \cos(\theta_B + \theta_T)\right)}, \\
& G(x_0, y_0) = \exp \left(- \frac{x_0^2 \cos^2 \theta_T \sec^2 \theta_B + y_0^2}{w_0^2} \right), \\
& H(x_0, t_0) = e^{i \phi_0} \, \operatorname{erfc}\left( \frac{c \beta (t_0 - 2 i \sigma_t^2 \omega_0) - x_0 \beta \sec \theta_B \sin \theta_T + x_0 \tan \theta_B}{2 c \beta \sigma_t} \right) + \mathrm{\emph{C.C.}}
\end{aligned}
\end{equation}
\end{widetext}
where \emph{C.C.} refers to adding the complex conjugate. In Fig.\ \ref{fig:momentum_comparison}, these analytic expressions are compared to a numerical integral of the forces. As before, even for the case of a few cycle pulse, the agreement is very good.

In order to produce the expressions in Eq.\ \ref{dpz}-\ref{dpx}, we need to make another approximation. Under the narrow spectrum approximation  $\sigma_t \omega_0 \gg 1$ that we have already made, and also when $t_0 \ll \sigma_t$, we can simplify further using the asymptotic form of the error function,
\begin{equation}
e^{i \phi} \mathrm{erfc}(\kappa - i n) + \mathrm{\emph{C.C.}} \approx - \frac{2 e^{n^2} \sin(\phi + 2 \kappa  n)}{\sqrt{\pi} n}.
\end{equation}
In this approximation all dependence on $\sigma_t$ is removed, so the momentum kick no longer goes to zero for large $t_0$, and therefore should only be used for small deviations in time $t_0$.

The same asymptotic form was also used to produce the Taylor series in Eq.\ \ref{dE_symm}. If the full form had been used instead, the coefficient of $t^2$ would instead be,
\begin{equation}
\begin{aligned}
& \frac{\omega_0^2}{2} t^2 \rightarrow \frac{\omega_0^2}{2} C(\sigma_t) t^2, \\
\text{where} \quad & C(\sigma_t) = \left( \frac{\exp(\omega_0^2 \sigma_t^2)}{\sqrt{\pi} \omega_0 \sigma_t \mathrm{Erfi}(\omega_0 \sigma_t)}\right).
\end{aligned}
\end{equation}

\begin{figure}
    \centering
    \includegraphics[width=\linewidth]{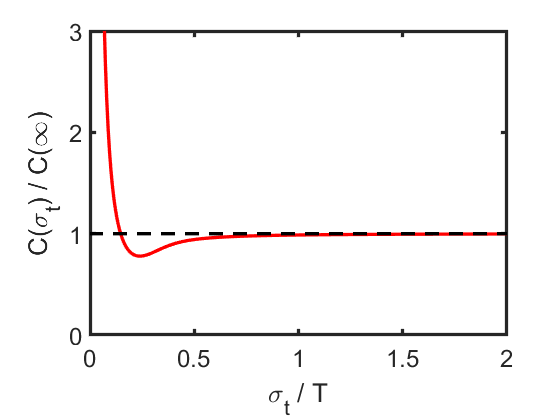}
    \caption{The relative strength of the $t^2$ coefficient in the Taylor expansion of $\Delta p_z$ as a function of the THz pulse length. Notably, the dependence is negligible for pulse lengths greater than the THz period $T$. This is the primary reason why the effectiveness of the THz monochromator has almost no dependence on the THz pulse length.}
    \label{fig:coeff}
\end{figure}

As shown in Fig.\ \ref{fig:coeff}, for any reasonable THz pulse length, there is an extremely weak dependence on $\sigma_t$, since the function $C(\sigma_t) \to 1$ for $\sigma_t \gtrsim 1/f$. This is the primary reason why the effectiveness of the monochromator has almost no dependence on the THz pulse length.

\section{THz Pulse Parameters} \label{Appendix_parameters}

\begin{figure}
    \centering
    \includegraphics[width=\linewidth]{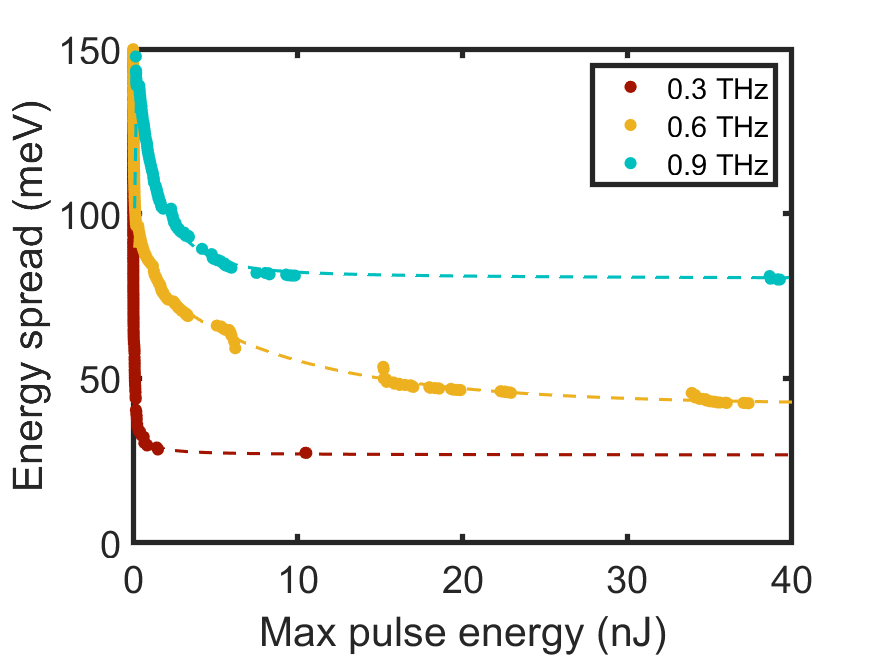}
    \caption{Multi-objective optimization fronts for two pulses at three representative frequencies.}
    \label{fig:2pulseFronts}
\end{figure}

\begin{figure*}
    \centering
    \includegraphics[width=\linewidth]{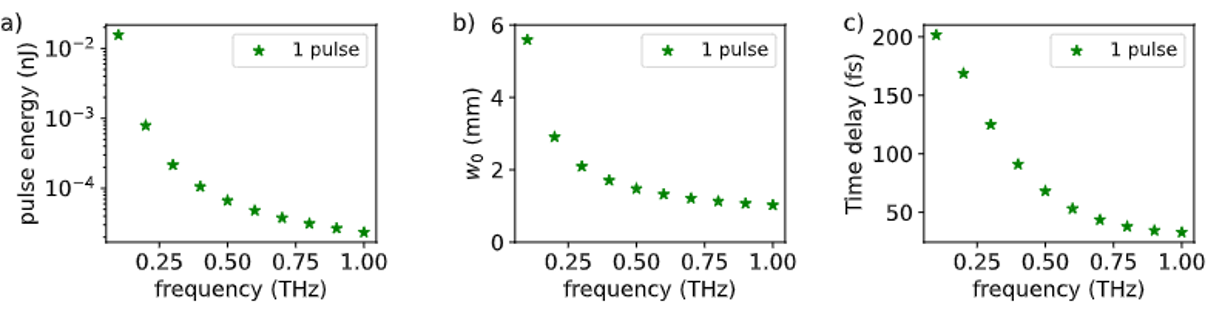}
    \caption{Optimal pulse parameters as a function of frequency for the single-pulse solutions shown in Fig.~\ref{fig:best_sigE}.}
    \label{fig:1pulseParams}
\end{figure*}

\begin{table*}[t]
\centering
\caption{Experimental and beamline parameters for three THz frequencies}
\label{tab:thz_parameters_singlecol}
\resizebox{\columnwidth}{!}{%
\begin{tabular}{
l
S[table-format=1.2]
S[table-format=1.2]
S[table-format=1.2]
l
}
\toprule
Name & {0.3 THz} & {0.6 THz} & {0.9 THz} & Units \\
\midrule
Pulse 1: Energy           & 0.16 & 2.25 & 5.14 & nJ \\
Pulse 1: Focal width, rms & 0.64 & 0.81 & 0.69 & mm \\
Pulse 1: Delay            & -1.27 & -0.41 & -0.92 & ps \\
Pulse 2: Energy           & 0.75 & 3.33 & 4.55 & nJ \\
Pulse 2: Focal width, rms & 0.81 & 0.83 & 0.71 & mm \\
Pulse 2: Delay            & 1.94 & 1.21 & 0.18 & ps \\
Pulse duration, rms       & 3.33 & 1.66 & 1.11 & ps \\
Gun voltage               & 60   & 60   & 60   & kV \\
Cathode--anode gap        & 1    & 1    & 1    & cm \\
Anode diameter            & 3    & 3    & 3    & mm \\
Solenoid diameter         & 1    & 1    & 1    & inch \\
Solenoid length           & 0.5  & 0.5  & 0.5  & inch \\
Solenoid strength         & 46   & 46   & 46   & kA-turns \\
Solenoid z position       & 0.25 & 0.25 & 0.25 & m \\
\bottomrule
\end{tabular}}
\end{table*}

The optimal parameters that yield the smallest energy spread at each frequency were found using a multi-variable numerical optimizer. The one-pulse configuration has a single solution, and follows a distinct trend in frequency (Fig.~\ref{fig:1pulseParams}). However, the two-pulse configuration is under-constrained, and there is a trade-off between each pulse parameter and the final energy spread. The primary practical constraint on the THz monochromator is the pulse energy, as it can be difficult to obtain large THz pulse energies by optical rectification. To address this we investigated the trade-off between energy spread and maximum pulse energy (between the two pulses, which are typically similar in magnitude) using a multi-objective genetic optimization algorithm. The resulting fronts that represent the compromise between pulse energy and $\sigma_E$ are shown in figure 12. The optimizer can always achieve a negligibly smaller energy spread by increasing pulse energy, but for all frequencies, reaches an asymptote in the range of tens of nanojoules. In Fig.~\ref{fig:best_sigE} we have plotted the asymptotic value of $\sigma_E$ at each frequency.

\begin{figure}[h!]
    \centering
    \includegraphics[width=\linewidth]{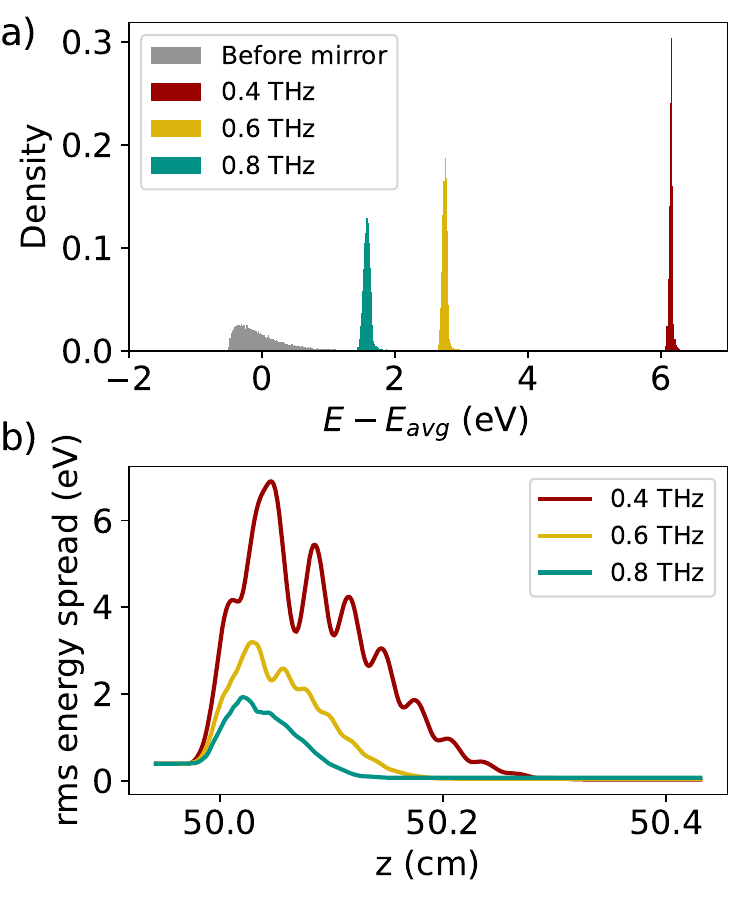}
    \caption{Summary results of particle tracking through the THz monochromator with two pulses of frequency 0.4, 0.6, and 0.8 THz. The electron beam has initial size $\sigma_x$ = 1 $\mu$m and duration 30 fs ($\sigma_t = 8.6$ fs) at the cathode. $\boldsymbol{a)}$  Histogram of particle energies before and after the THz interaction.  $\boldsymbol{b)}$ Energy spread vs propagation distance, after the THz interaction point. }
    \label{fig:z_vs_sigE}
\end{figure}

\clearpage
\nocite{}

\bibliography{Cecilia}

\end{document}